\documentstyle[12pt]{article}
\begin{document} 

\title{Solutions of Quantum Gravity Coupled to the Scalar Field} 
\author{A.\ B\l{}aut\thanks{e-mail address 
ablaut@ift.uni.wroc.pl}\ \ and J.\ Kowalski--Glikman\thanks{e-mail 
address 
jurekk@ift.uni.wroc.pl}\\ 
Institute for Theoretical Physics\\ 
University of Wroc\l{}aw\\ 
Pl.\ Maxa Borna 9\\ 
Pl--50-204 Wroc\l{}aw, Poland} 
\maketitle 
 
\begin{abstract} 
We consider the Wheeler--De Witt equation for canonical quantum 
gravity
coupled to massless scalar field. After regularizing and renormalizing
this equation, we find a one-parameter class of its solutions.
\end{abstract} 
\vspace{12pt} 
PACS number 04.60 Ds 
\clearpage

The term ``quantum gravity'' plays a quite magic role in modern 
theoretical
physics. All the major problem troubling our understanding of the 
nature
in extreme conditions, like the cosmological constant problem, the
problem of the Big Bang (or the problem of initial conditions for
inflationary cosmological scenario), the problem of black hole
evaporation, etc.\ are all supposed to disappear, or become finally
resolved once this Holy Grail of modern physics becomes understood. In
recent years, thanks to the rapid developments in many approaches to 
the
problem \cite{reviews} (superstrings theory, loop variables, lattice 
quantum gravity,
and functional approach): we see some shapes of the ultimate theory 
(or
theories) that, in principle, could be checked against some 
experimental
(mostly cosmological) data and general theoretical understanding of 
the
nature.

In the recent papers \cite{JK}, \cite{JA} we investigated a scheme of
quantization of pure gravitational field based on the traditional
Wheeler--De Witt approach. The results of these papers can be 
summarized
as follows:

\begin{enumerate}
\item We choose the standard ordering of the diffeomorphism 
constraint,
to wit
\begin{equation} 
{\cal D}_a=\nabla_b\, \pi^{ab}. \label{diffconstr}
\end{equation}
This makes the states being (compact) space integrals of scalar
densities manifestly diff--invariant.
\item As for the hamiltonian constraint (Wheeler--De Witt operator), 
we
write it in the regularized form with an additional term
responsible for ordering ambiguity:
$$ 
{\cal H}(x)=\kappa^2\int\, dx'\, G_{abcd}(x')K(x,x';t)
\frac{\delta}{\delta h_{ab}(x)}\frac{\delta}{\delta h_{cd}(x')} + 
$$ 
\begin{equation} 
+\kappa^2 \left(\lambda_{1} h_{ab}+ (\lambda_{2} h_{ab}R 
+ \lambda_{3} R_{ab})\right)(x) 
\frac{\delta}{\delta h_{ab}(x)}+ 
 \frac1{\kappa^2}\sqrt h 
(R +2\Lambda),\label{qhamf} 
\end{equation} 
where 
$$ 
K(x,x';t)=\frac{\exp\left(-\frac1{4t} 
N_{ab}(x)(x-x')^a(x-x')^b\right)}{4\pi t^{3/2}}\bar{K}(x,t) 
$$ 
with $\bar{K}(x,t)$ being a power series in $t$ vanishing at $t=0$.
Using the fact that $t$ has dimension $m^{-2}$ we make the following
expansion for $\bar{K}$ and $N_{ab}$ 
\begin{equation} 
\bar{K}(x,t)=1+a_0tR+(a_1 R^2 + b_1 R_{ab}R^{ab}) + \ldots, 
\label{kexp}
\end{equation}
$$
N_{ab}(x) = h_{ab} + 2t(A_0 R_{ab} + B_0 h_{ab}R) + $$
\begin{equation}
+2 t^2( B_1 R_a^cR_{cb} + A_1 R_{ab} R + C_1 h_{ab} R^2 + D_1 
h_{ab}R^{cd}R_{cd}) +
 \ldots, 
\label{nexp}
\end{equation} 
where \ldots denote the higher order terms. $a$, $b$,  
$A$, $B$, $C$ $\lambda_{1}$, $\lambda_{2}$, $\lambda_{3}$ are the 
free 
parameters which in the paper \cite{JA}  we fixed by the requirement 
that the 
constraints algebra is anomaly-free and that the WDW operator has a 
maximal 
number of possible solutions.
\item We then solve the resulting equation taking the ansatz that the
wavefunction of the universe is a function of the volume of the space
${\cal V}= \int \sqrt h\, dx$ and the average curvature ${\cal R}=
\int \sqrt h R\, dx$.
\item Given the wavefunctions spanning
(part of) the space of solutions,  we seek their interpretation making
use of the quantum potential approach to quantum theory \cite{qpot}.
\end{enumerate}
\vspace{12pt}

In the present paper we basically repeat
these steps\footnote{We do not address the issue of anomaly, though. 
The
reason is that the experience with purely gravitational case shows 
that
if solutions exist, then in their neighborhood, at least, there is no
anomaly.}  in the case of the
quantum gravity coupled to the  scalar field. The
necessary modifications consist of additional terms in diffeomorphism
and hamiltonian constraints. In the former case we again use the 
natural
ordering guaranteeing that space integrals of scalar densities are
diff-invariant. As for the Wheeler--De Witt equation, we add to
(\ref{qhamf}) a regularized
and arbitrarily ordered kinetic term
\begin{equation}
\int dx'\, W(x,x',s)\frac{\delta}{\delta \phi(x)}\frac{\delta}{\delta
\phi(x')} + S(x)\frac{\delta}{\delta \phi(x)}, \label{phikin}
\end{equation}
with $W(x,x',s)$ being constructed basically by replacing powers of 
$R_{ab}$ in
$K(x,x't)$ with powers of $\nabla_a\phi\nabla_b\phi$ (with $\phi$
in units of the bare Planck mass),
along with the potential terms
\begin{equation}
- \sqrt h  \frac12 (\nabla \phi)^2 .
\label{phipot}
\end{equation}
In what follows we will consider only the massless scalar field. The 
reason
is that we are considering a theory which is supposed to be an 
ultimate
theory of the nature and thus we should keep the minimal possible 
number
of coupling constants. We hope that some solutions will describe the
massive scalar field. 

Let us turn now to the renormalization problem. Clearly, the 
action of the Wheeler-De Witt operator on a states being a 
function of integrated scalar densities gives as a result
a series containing both positive and negative powers of
$t$. To renormalize, we make use of the analytic 
continuation proposal of Mansfield \cite{Mansfield} which results with
replacing all positive powers of $t$ by zero and 
negative powers $t^{-n/2}$ by the renormalization constants
$\rho^{(n)}$. In what follows we will make use of the following
formulas for coincidence limit of the regulators.

$$
K(x,x)_{ren}=\rho^{(3)}+a_0 \rho^{(1)}R
$$ 
$$     
\left.\left( \Box_{x'}K(x,x')\right|_{x=x'} 
\right)_{ren}=-\frac{3}{2}\rho^{(5)}+(A_{0}+
3B_{0}-\frac{3}{2}a_{0})R\rho^{(3)}
$$
$$
+\left[ (A_{0}+3B_{0})a_{0}+A_{1}+3C_{1}-\frac{3}{2}a_{1}  \right] 
R^{2}\rho^{(1)}+(B_{1}+3D_1-\frac{3}{2}b_{1})Ric^{2}\rho^{(1)}
$$ 
$$ 
W(x,x)={\rho}^{(3)}+\bar{a}_{0}{\rho}^{(1)}\kappa^2(\nabla\phi)^2
$$
$$
\left.\left( \Box_{x'}W(x,x')\right|_{x=x'} 
\right)_{ren}=-\frac{3}{2}{\rho}^{(5)}+(\bar{A}_{0}+
3\bar{B}_{0}-
\frac{3}{2}\bar{a}_{0})\kappa^2(\nabla\phi)^{2}{\rho}^{(3)}
-\frac32\rho^{(1)}\bar e_1 \kappa^2\Box\phi\Box\phi +
$$
$$
\left[ (3\bar{B}_{0}+\bar{A}_{0})\bar{a}_{0}+3\bar{C}_{1}+\bar A_{1}
-\frac{3}{2}\bar{a}_{1}  \right] \kappa^4(\nabla\phi)^{4}{\rho}^{(1)}+
(3\bar{D}_{1}-\frac{3}{2}\bar{b}_{1})\kappa^4(\nabla_{a}\phi\nabla_{b}
\phi)^{2}
{\rho}^{(1)}
$$
$$
S(x)=s_{1}\Box\phi
$$
We choose our wavefunction to be of the form
\begin{equation}
\label{psi}
\Psi=e^{\alpha{\cal F}}e^{\beta{\cal R}}e^{\gamma{\cal P}}
\end{equation}   
with
$$
 {\cal R}=\int\sqrt{h}R, \;
{\cal F}=\int\sqrt{h}f(\phi), \;
{\cal P}=\int\sqrt{h}(\nabla\phi)^2,
$$
where $f(\phi)$ is at this stage an arbitrary function.
Clearly this wavefunction is diffeomorphism-invariant. Substituting
this form to the WDW equation (\ref{qhamf}), we obtain a number of
independent terms whose coefficients are to vanish.
These coefficients, in turn, depend on  $\alpha, \beta,\gamma$
and the parameters of the regulator which will become fixed in the 
process of solving of the equation. It should be observed
that since every regulator corresponds, in principle, to
a distinct quantum theory we effectively consider only those 
theories of quantum geometry that possess as a solutions 
the class of states of the form (\ref{psi}).
Now the procedure is rather straightforward. We present it 
term-by-term:

\begin{enumerate}

\item From the $R_{ab}\nabla_{a}\phi\nabla^{b}\phi$ term we find 
\begin{equation}
\beta=\frac{\lambda_{3}}{2}
\end{equation}      

\item From terms involving $\Box\phi\Box\phi$ we get
\begin{equation}
\gamma=\frac{3}{4}\rho^{(1)}\kappa^{2}\bar{e}_{1}
\end{equation}

\item All other terms involving four derivatives give relations which 
can be 
solved to express higher order coefficients in terms of the lower
order ones. Since these coefficients do not appear in what follow
we will not present the resulting equations explicitely. The only
exeption is the coefficient of $R(\nabla\phi)^2$ term which gives
\begin{equation}
\label{item3}
4\lambda_{2}+\lambda_{3}+7a_{0}\rho^{(1)}=0
\end{equation}
\end{enumerate}

Consider now terms containing $R$ and $Rf$. Since $f$
is an arbitrary function,  the coefficients multiplying
these terms must vanish separately. 

We find again condition for $\lambda_{2}$ and $\lambda_{3}$
which together with (\ref{item3}) gives
\begin{eqnarray}
\lambda_{2}&=&-\frac{7}{6}a_{0}\rho^{(1)}\\
\lambda_{3}&=&-\frac{35}{24}a_{0}\rho^{(1)}
\end{eqnarray}
and the equation 
\begin{eqnarray}
\frac{1}{\kappa^{4}}&=&-\frac{35}{48}a_{0}\rho^{(1)}\rho^{(3)}\left(
\frac{7}{8}+A_{0}+3B_{0}-\frac{1}{2}a_{0}\right)\nonumber\\
&&+\frac{35}{48}a_{0}\rho^{(1)}\lambda_{1}
\end{eqnarray}

Next let us turn to terms containing $(\nabla\phi)^2$. One gets the
following differential equation
\begin{equation}
f''-\Omega^2f=F
\end{equation}
where 
\begin{eqnarray}             
\Omega^2&=&\frac{1}{2}\frac{\gamma}{\bar{a_{0}}\rho^{(1)}}=
\frac{3\bar{e}_{1}\kappa^{2}}{8\bar{a}_{0}}\\
F&=&\frac{3\bar{e}_{1}\kappa^{2}}{2\bar{a}_{0}\alpha}\left[
-\frac{7}{8}\rho^{(3)}-\frac{1}{2}\lambda_{1}+(\bar{A}_{0}+3\bar{B}_{0}
-\frac{3}{2}\bar{a}_{0})\rho^{(3)}\right]
\end{eqnarray}

Thus
\begin{equation}
f=Ae^{\Omega\phi}+Be^{\Omega\phi}-\frac{F}{\Omega^{2}}=f_{0}(\phi)
-\frac{F}{\Omega^{2}}
\end{equation}
To complete solving WDW equation we write down all the remaining terms 
in the
form of a single equation. 

It reads
\begin{eqnarray}
&&\frac{3}{8}\alpha^{2}\kappa^{2}f^{2}-\frac{1}{2}\alpha^{2}f'^{2}
-\frac{3}{2}\beta\kappa^{2}\rho^{(5)}+
\frac{21}{8}\alpha\kappa^{2}f\rho^{(3)}\nonumber\\
&&-\frac{3}{2}\alpha\lambda_{1}f\kappa^{2}-\frac{1}{2}\alpha 
f''\rho^{(3)}
-\frac{3}{2}\gamma\rho^{(5)}-\frac{2\Lambda}{\kappa^{2}}=0
\end{eqnarray} 
Consider first the part nonlinear in $f_{0}$. We get
\begin{equation}
\frac{3}{8}\kappa^{2}f_{0}^{2}-\frac{1}{2}f'^{2}_{0}=0
\end{equation}
from which we obtain the condition
\begin{equation}
\Omega^{2}=\frac{3}{4}\kappa^2
\end{equation}
and the additional constant piece $\frac{3}{2}\alpha^{2}\kappa^{2}AB$.
Now we consider terms linear in $f_{0}$ from which we find an 
expression
for $\alpha$.

The last equation gives an expression of the form
\begin{equation}
\alpha^{2}AB=const
\end{equation}
which gives
\begin{equation}
\alpha B=\frac{C}{\alpha A}
\end{equation}
\vspace{12pt}

Thus our final result for the wavefunction (after rescaling
$A$ by $\alpha$) is of the form
\begin{eqnarray}
{\Psi}_{A}&=&\exp{\left[\left( 7\rho^{(3)}-8(\bar{A}_{0}
+3\bar{B}_{0}-\frac{3}{2}\bar{a}_{0})\rho^{(3)} +4\lambda_{1}\right)
{\cal V}\right]}
\exp{\left(\frac{3}{4}\rho^{(1)}\kappa^{2}\bar{e}_{1}{\cal 
P}\right)}\nonumber\\
&&\exp{\left(-\frac{35}{48}a_{0}\rho^{(1)}{\cal 
R}\right)}
\exp{\int d^{3}x\left( 
Ae^{\frac{\sqrt{3}}{2}\kappa\phi} 
+\frac{C}{A}e^{-\frac{\sqrt{3}}{2}\kappa\phi}  
\right)}\label{wff}\end{eqnarray}with ${\cal V}=\int\, d^3x \sqrt h$ being 
the volume of spaces.

This equation completes our construction of solutions
of WDW equation for quantum gravity
coupled to massless scalar field.
\newline

Let us conclude this paper with some comments.

\begin{enumerate}
\item We found a class of solutions parametrized by an arbitrary 
complex constant
$A$. Since the system is linear, any complex combination of solutions 
is
a solution again and we may seek interpretation of such solutions 
making
use of the quantum potential approach (as in \cite{JA}.)
\item The same procedure will make it possible to discuss in full the
problem of renormalization. Our wavefunction (\ref{wff}) depends on 
both
bare coupling parameters ($\kappa$ and $\Lambda$) and the arbitrary
renormalization parameters $\rho$. In quantum potential approach one
obtains the corresponding classical evolution modified by additional
quantum terms. Thus it is possible to relate these both sets of
parameters to the physical gravitational and cosmological constants 
and
to derive the low energy mass of the scalar field.
\item As a rule the quantum corrections in the quantum potential
approach is nonlocal (depending on  the volume of space ${\cal V}$,
average curvature ${\cal R}$, etc.) This means that the effective
constants of the theory depend on global properties of space. It is
likely that this approach may shed some light on the problem of the
origin of the inflaton field and the cosmological constant problem. 
All
these questions will be discussed in the forthcoming paper.

\end{enumerate}

\end{document}